# Link Prediction in Complex Networks by Multi Degree Preferential-Attachment Indices

Ke Hu[1,*], Ju Xiang[2], Wanchun Yang[3], Xiaoke Xu[4] and Yi Tang[1]

[1]*Department of Physics, Xiangtan University, Xiangtan 411105, Hunan, China*
[2]*First Aeronautical College of Air Force, Xinyang 464000, Henan, China*
[3]*College of Information Engineering, Xiangtan University, Xiangtan 411105, Hunan, China*
[4]*College of Information and Communication Engineering, Dalian Nationalities University, Dalian 116605, China*

*E-mail: huke1998@yahoo.cn

**Abstract:** In principle, the rules of links formation of a network model can be considered as a kind of link prediction algorithm. By revisiting the preferential attachment mechanism for generating a scale-free network, here we propose a class of preferential attachment indices which are different from the previous one. Traditionally, the preferential attachment index is defined by the product of the related nodes' degrees, while the new indices will define the similarity score of a pair of nodes by either the maximum in the two nodes' degrees or the summarization of their degrees. Extensive experiments are carried out on fourteen real-world networks. Compared with the traditional preferential attachment index, the new ones, especially the degree-summarization similarity index, can provide more accurate prediction on most of the networks. Due to the improved prediction accuracy and low computational complexity, these proposed preferential attachment indices may be of help to provide an instruction for mining unknown links in incomplete networks.

## 1. Introduction

Link prediction, aiming at estimating the likelihood of the existence of a link between two agents based on observed links and the attributes of agents [1,2], is revealing a rich subject with a wide scope of applications [3-7]. In the practical aspect, link prediction can provide significant instruction for mining missing interactions in incomplete networks [8-10]. Usually, the identification of missing links in experiment requires the time-consuming and expensive examination of a great number of possible connections. This sometimes is unacceptable in practice. In this case, the prioritization of missing links may be of help to the identification of missing links, since it can largely reduce both the time consume and experimental cost [3]. In the aspect of theory, the link prediction algorithms are beneficial for the understanding of the evolution of real networks. Accurate prediction of links in a network may provide some important clues or evidences about the underlying mechanism that drives its evolution [7]. With the help of link prediction algorithms, one may construct a proper evaluation system to evaluate the evolving mechanism for given networks [6]. Starting from these facts, the link prediction algorithms can be extended to broader applications. Recently, some prediction algorithms have been successfully applied to the classification problem in partially labeled networks [11,12] as well as the identification of the spurious links resulting from inaccurate information in the data [5].

In the past few years, a large number of methods have been developed for predicting unknown links in complex networks [2-4, 13-21]. Generally, these methods are designed respectively based on two types of different information: the external information besides the network topology such as the unit attributes [13-16] and the information from the network structure [17-21]. In view of the difficulty of the external information acquisition, the information of network topology is usually preferable to the link prediction problem. Among these network-based methods, one of the simplest and effective algorithms is that based on the preferential attachment (PA) index [17]. It is motivated by the popular preferential attachment (PA) mechanism in evolving scale-free network models [22]. Originally, this mechanism just describes interactions between the newly added node and the old ones [22]. Actually, it can also describe the interaction between two old nodes [23-25]. Now, the traditional way to express this interaction is by the product form of related nodes' degrees, i.e., the pairwise interaction between nodes $i$ and $j$ is proportional to $k_i k_j$ (degree-product form). In some real networks, the formation of new links between two old nodes indeed follows the degree product form [26-29]. Benefited from the long-held and partially proved assumption of degree product form, Zhou et al [17] proposed the *Degree-Product Preferential Attachment* (DPPA) index to perform the link prediction in networks. Although the DPPA index is rooted in the popular PA mechanism of network evolution and have a good physical basic, it cann't give good prediction for most sample networks [17]. In this paper, by analyzing the PA mechanism in detail, we find that the degree product form is not the most suitable one for the link prediction problem, and therefore we develop a class of new PA indices. Extensive experiments on fourteen real networks demonstrate that the new PA indices can give more accurate prediction than the traditional DPPA index in most sample networks.

## 2. Datasets and Method
2.1 Datasets
In this paper, fourteen networks that are draw from different fields are considered in our experiment. These networks are simply described as follows:

1) Collaboration network in Computational Geometry (CCG) [30]: It is an author-collaboration network in which nodes represent authors, and two authors are linked with a link if they wrote a common work (paper, book, etc.). Multiple links between two nodes represent multiple joint works they wrote. In this paper, we consider only unweighted network, and thus the multiple links between a pair of authors are replaced with a single link.

2) Internet (INT) [31]: The Internet can be decomposed into subnetworks that are under separate administrative authorities. Here, we consider the Internet at the level of autonomous systems (the inter-domain level) where each domain is represented by a single node and each link is an inter-domain interconnection.

3) USAir [32]: The network of the US air transportation system in which nodes and links represent airports and airlines respectively.

4) Protein-Protein Interaction network (PPI) [33]: A protein-protein interaction network in budding yeast with nodes representing proteins and links corresponding to the interactions among proteins.

5) FoodWeb (FW) [34]: A network of foodweb in Florida Bay during wet season. Each species is represented as a node of the network, and a link is placed between two species whenever one of them feeds on the other.

6) NetScience (NS) [35]: A network of coauthorships between scientists who are themselves publishing on the topic of networks.

7) The network of Common Adjective and Noun adjacencies (CAN) [35]: It is the network of common adjective and noun adjacencies for the novel "David Copperfield" by Charles Dickens. Nodes represent the most commonly occurring adjectives and nouns in the book. Links connect any pair of words that occur in adjacent position in the text of the book.

8) World Wide Web (WWW) [36]: A large directed graph whose nodes are documents from University of Notre Dame (domain nd.edu) and whose edges are links (URLs) that point from one document to another.

9) Gnutella Peer-to-peer network (GP) [37]: A sequence of snapshots of the Gnutella peer-to-peer file sharing network from August 2002. It is pure directed network and nodes represent hosts in the Gnutella network topology and links represent connections between the Gnutella hosts.

10) Enron Email network (EE) [38]: Enron email communication network covers all the email communication within a dataset of around half million emails. Nodes of the network are email addresses and if an address $x$ sent at least one email to address $y$, the graph generates a directed link from $x$ to $y$. Note that non-Enron email addresses act as sinks and sources in the network as we only observe their communication with the Enron email addresses.

11) Epinions Social network (ES) [39]: This is a who-trust-whom online social network of the general consumer review site—Epinions.com. Members of the site can decide whether to "trust" each other. All the trust relationships interact and form the Web of Trust which is then combined with review ratings to determine which reviews are shown to the user.

12) Slashdot Social network(SS) [40]: Slashdot is a technology-related news website known for its specific user community. The website features user-submitted and editor-evaluated current primarily technology oriented news. In 2002 Slashdot introduced the Slashdot Zoo feature which allows users to tag each other as friends or foes. The network contains friend/foe links between the users of Slashdot.

13) E-mail (EM) [41]: The e-mail network studied here is the email network of University at Rovirai Virgili (URV) in Tarragona, Spain, and is built regarding each email address as a node and linking two nodes if there is an email communication between them.

14) Power Grid (PG) [42]: An electrical power grid of the western US, with nodes representing generators, transformers and substations, and links corresponding to the high voltage transmission lines between them.

It should be mentioned that some of the above networks consist of many separated components but the sizes of the largest connected components relative to the whole networks are still very large, most of the nodes in a network belong to the same large connected component, so we will only consider the giant connected component in the networks. In addition, some networks mentioned above are directed and/or weighted. In this paper, we focus on the prediction for undirected and unweighted links, i.e., we don't consider the effects of direction and weights on link prediction. Thus the networks will be treated as undirected and unweighted networks. The basic topological features of the fourteen real-world networks are summarized in table 1.

Table 1. The basic topological features of fourteen example networks. $N$ and $M$ are the total numbers of nodes and links, respectively. $N_C$ denotes the size of the giant component. For example, the entry 3621/2091 in the first line means that the network has 2091 components and the giant component consists of 3621 nodes. $M_C$ is the number of links belonging to the giant component. $e$ is the network efficiency [43], defined by $e = \frac{1}{N(N-1)} \sum_{x \neq y \in E} \frac{1}{d_{xy}}$, where $d_{xy}$ is the shortest path length between $x$ and $y$, and $d_{xy} = +\infty$ if x and y are in two different components. $C$ and $r$ are the clustering coefficient [42] and assortative coefficient [44], respectively. $H$ is the degree heterogeneity, defined as $H = \langle k^2 \rangle / \langle k \rangle^2$, where $\langle k \rangle$ denotes the average degree [17].

| Networks | $N$ | $M$ | $N_C$ | $M_C$ | $e$ | $C$ | $r$ | $H$ |
|---|---|---|---|---|---|---|---|---|
| CCG | 7343 | 11898 | 3621/2091 | 9461 | 0.0506 | 0.4075 | 0.2426 | 4.7062 |
| INT | 22963 | 48436 | 22963/1 | 48436 | 0.2757 | 0.2304 | -0.1984 | 61.9778 |
| USAir | 332 | 2126 | 332/1 | 2126 | 0.4059 | 0.6252 | -0.2079 | 3.4639 |
| PPI | 2361 | 7182 | 2224/161 | 6609 | 0.2183 | 0.2910 | 0.0587 | 2.7632 |
| FW | 128 | 2137 | 128/1 | 2106 | 0.6243 | 0.3346 | -0.1044 | 1.2307 |
| NS | 1461 | 2742 | 379/268 | 914 | 0.0163 | 0.6937 | 0.4616 | 1.8486 |
| CAN | 112 | 425 | 112/1 | 425 | 0.4420 | 0.1728 | -0.1293 | 1.8149 |
| WWW | 325729 | 1090108 | 325729/1 | 1090108 | 0.1535 | 0.2346 | -0.0534 | 41.9342 |
| GP | 62586 | 147892 | 62561/12 | 147878 | 0.1735 | 0.0055 | -0.0926 | 2.4552 |
| EE | 36692 | 183831 | 33696/1065 | 180811 | 0.2214 | 0.4970 | -0.1108 | 13.9796 |
| ES | 75879 | 405740 | 75877/2 | 405739 | 0.2445 | 0.1378 | -0.0406 | 17.1939 |
| SS | 82140 | 500481 | 82140/1 | 500481 | 0.2568 | 0.0588 | -0.0730 | 12.1529 |
| EM | 1133 | 5451 | 1133/1 | 5451 | 0.2999 | 0.2202 | 0.0782 | 1.9421 |
| PG | 4941 | 6594 | 4941/1 | 6594 | 0.0629 | 0.1071 | 0.0035 | 1.4503 |

## 2.2 Preferential Attachment similarity indices

Here we are still interesting in the preferential attachment similarity due to two significant reasons: one is that it requires the least information since it only depends on the degrees of related nodes and the another is that it has a good physics basic since it is originated from the popular PA evolving mechanism of generating the scale-free network [22]. Motivated by this PA mechanism, the traditional similarity index, i.e., the DPPA index, has been defined by the degree-product form,

$$s_{xy} = k_x \times k_y, \quad (1)$$

where $k_x$ and $k_y$ are the degrees of nodes $x$ and $y$ respectively. This index has been used to quantify the functional significance of links subject to various network-based dynamics, such as percolation [45], synchronization [46] and transportation [47]. For the application of the PA mechanism to the link prediction, one may naturally think of the above degree-product form, but compared with almost all other node-similarity indices, the degree-product DPPA index is poor in the link prediction. So we argue that the degree-product form is not most reasonable for the link prediction in real networks, although it perhaps can represent the interaction between two nodes in some cases.

In principle the rules of the additions of links can be considered as a kind of link prediction algorithm, which thus builds a bridge between the link prediction and the mechanism of evolving models [6]. Because a proper link prediction can give evidence to some underlying mechanisms that drive the network evolution; inversely, the legitimate mechanism of network evolution can also provide significant clues to design a proper predictive algorithm. Now, let's revisit the preferential attachment mechanism. It is well known that the preferential attachment generally appears at two levels [29]. We will give two new preferential attachment indices corresponding to the two levels.

*i) Links appear between the newly added node and the old ones:* a new node $x$ is added along with several new links each of which is randomly attached to an existing node $y$ by the degree preferential attachment probability,

$$\Pi_{x \to y}^{new-old} = \frac{k_y}{\sum_j k_j}. \qquad (2)$$

Clearly, this probability is proportional to the degree of the existing node $y$, independent of the degree of the newly added node. Note that the degree preferential attachment probability is a conditional probability that states the link formation from certain nodes (the newly added nodes) to old nodes in evolving network model, while not the preferential attachment probability of link formation between any pair of nodes. However, the DPPA index is designed to describe the similarity between any pair of nodes, and thus do not correspond to the preferential mechanism completely. In addition, because the degrees of the newly added nodes (young nodes) are generally smaller than the old nodes, we may distinguish between the new nodes and the old nodes by their degrees at the moment of the link formation. Based on the above analysis, a complete counterpart to this PA mechanism and a more reasonable form of the PA index in this case can be designed as,

$$s_{xy} = Max\left[k_x, k_y\right]. \qquad (3)$$

For convenience, we name this index as the *High-degree node Determine Preferential Attachment* (HDPA) index.

*ii) Internal links appear between two old nodes:* Except for the above new link addition from the new nodes, another way can not be neglected: a large number of new links may appear between old nodes as the network evolving. Such internal links are often also subject to preferential attachment. Following the above link formation from new nodes to old nodes, a link between two old nodes $x$ and $y$ may be constructed by three ways: (a) link formation from node $x$ to node $y$, (b) link formation from node $x$ to node $y$ or (c) both. So the probabilities of link formation corresponding to the three cases are $\Pi_{x \to y} = \frac{k_y}{\sum_j k_j}$, $\Pi_{y \to x} = \frac{k_x}{\sum_j k_j}$ and $\Pi_{x \leftrightarrow y} = \frac{k_x k_y}{\sum_j k_j \sum_j k_j}$, respectively. And then the corresponding similarity scores for link prediction can be wrote as: $s_{xy}^{x \to y} = k_y$, $s_{xy}^{y \to x} = k_x$, and $s_{xy}^{x \leftrightarrow y} = k_x k_y$. In order to integrate the three indices, we can rescale the first two indices as $s_{xy}^{x \rightarrow y} = k_y^2$ and $s_{xy}^{y \rightarrow x} = k_x^2$, since the rescaling doesn't change the order of the similarity scores of links and make their scores comparable. Then we integrate them into one by simple weighted summarization:

$$s_{xy} = ak_x^2 + bk_y^2 + ck_x k_y, \tag{4}$$

where *a*, *b* and *c* is three adjustable parameter to control the relative contributions of the three measures to the integrated one. Consider the symmetry of link formation from node *x* to *y* and from *y* to *x* and without lack of generality, we can set $a = b = \varepsilon$ and $c = 1$, and obtain a *General Internal-Links Preferential Attachment* (GILPA) index:

$$s_{xy} = \varepsilon k_x^2 + \varepsilon k_y^2 + k_x k_y. \tag{5}$$

Clearly, the index reduces to the traditional DPPA index when $\varepsilon = 0$. Particularly, when $\varepsilon = 0.5$, we can obtain a very simple similarity measure:

$$s_{xy} = k_x + k_y. \tag{6}$$

We name it as the *Degree-Summation Preferential Attachment* (DSPA) index. Another limiting case is $\varepsilon \rightarrow +\infty$, which corresponds to a *Degree-Squared-Summation Preferential Attachment* (DSSPA) index:

$$s_{xy} = k_x^2 + k_y^2. \tag{7}$$

Of course, given a network, one can tune $\varepsilon$ to find its optimal value corresponding to the highest accuracy, however this optimal value is different for different networks, and a parameter-dependent measure is less practical in dealing with huge-size networks since the tuning process may take much time.

In the above discussion, we have divided the link formations into two types according to the rules of their addition and present two new similarity indices to predict these links. The two indices are clearly different from the DPPA index. This is because the PA probabilities that they depends on (which is not the degree-product from) are different from that between two old nodes in random networks with given degree distribution. However, which index is best? We need the experimental confirmation.

### 2.3 Evaluation metrics

Let *G*(*V*, *E*) be a simple undirected and unweight network, which is described by the sets of nodes *V* and links *E*. Multiple links and self-connections are excluded from *E*. Every algorithm referred in this paper will assign a similarity matrix *S* whose real entry $s_{xy}$ expresses how similar node *x* is to node *y*: we say that $s_{xy}$ is their similarity score. For each pair of nodes *x* and *y* (x, y∈*V*), $s_{xy}=s_{yx}$ since the networks are undirected. All the nonexistent links are sorted in decreasing order according to their similarity scores, and the links at the top are most likely to exist. To quantify the prediction accuracy, the set of the observed links *E* is randomly divided into two parts: the training set $E^T$ and the probe set $E^P$. The training set is treated as known information, while the probe set will be predicted and no information in this set is allowed to be used for prediction. Clearly, $E = E^T \cup E^P$ and $E^T \cap E^P = \varnothing$. In this paper, the training set always contains 90% of links and naturally the remaining 10% of links constitute the probe set.

The prediction quality is then evaluated by the standard metric, the Area Under the receiver operating characteristic Curve (AUC) [48]. In the present case, this metric can be interpreted as the probability that a randomly chosen missing link (a link in $E^P$) is given a higher similarity score than a randomly chosen nonexistent link (a link in $U$-$E^T$, where U denotes the universal set). In the implementation, among $n$ independent comparisons, if there are $n'$ occurrences of the missing link having a higher score and $n''$ occurrences of the missing link and nonexistent link having the same score, we define the accuracy as

$$AUC = \frac{n'+0.5n''}{n}. \tag{8}$$

If all the scores are generated from an independent and identical distribution, AUC should be about 0.5. Therefore, the degree to which the value exceeds 0.5 indicates how much better the algorithm performs than pure chance.

## 3 Results

Table 2. Accuracies of algorithms, measured by the area under the ROC curve. Each number is obtained by averaging over 10 implementations with independently random partitions of testing set and probe set.

| Networks | DPPA | HDPA | DSSPA | DSPA | GILPA ($\varepsilon_{optimal}$) |
|---|---|---|---|---|---|
| CCG | 0.7222 | 0.7681 | 0.7684 | 0.7707 | 0.7720($\varepsilon$=0.67) |
| INT | 0.7722 | 0.9560 | 0.9531 | 0.9481 | 0.9531($\varepsilon$>10.0) |
| USAir | 0.8789 | 0.8537 | 0.8670 | 0.8800 | 0.8861($\varepsilon$=0.28) |
| PPI | 0.7808 | 0.8130 | 0.8202 | 0.8285 | 0.8302($\varepsilon$=0.23) |
| FW | 0.6912 | 0.6572 | 0.6733 | 0.6913 | 0.6913($\varepsilon$=0.50) |
| NS | 0.6368 | 0.6458 | 0.6499 | 0.6556 | 0.6574($\varepsilon$=0.42) |
| CAN | 0.7392 | 0.7113 | 0.7120 | 0.7548 | 0.7548($\varepsilon$=0.50) |
| WWW | 0.8167 | 0.9157 | 0.9218 | 0.9252 | 0.9262($\varepsilon$=0.24) |
| GP | 0.7171 | 0.7903 | 0.8116 | 0.8244 | 0.8269($\varepsilon$=0.25) |
| EE | 0.8792 | 0.9033 | 0.9089 | 0.9142 | 0.9246($\varepsilon$=0.02) |
| ES | 0.8849 | 0.9333 | 0.9344 | 0.9368 | 0.9374($\varepsilon$=0.21) |
| SS | 0.9007 | 0.9234 | 0.9283 | 0.9305 | 0.9360($\varepsilon$=0.07) |
| EM | 0.7820 | 0.7447 | 0.7601 | 0.7775 | 0.7872($\varepsilon$=0.04) |
| PG | 0.4421 | 0.5428 | 0.5242 | 0.5028 | 0.5242($\varepsilon$>10.0) |

In table 2, we give the numerical results of these similarity indices in the fourteen sample networks. As is not unexpected, for different networks, these PA indices give distinct link-prediction accuracies. Moreover, the performances of these indices strongly depend on the degree heterogeneity. If all the nodes in a given network have pretty much the same degree, corresponding to a very small $H$, then they will give relatively bad predictions. On the contrary, larger is the degree heterogeneity, higher are their accuracies, which can be seen in Fig. 1(a). This indicates that the degree heterogeneity has an important implication to the PA mechanism. Perhaps the degree heterogeneity is indeed originated from the PA mechanism, or more specifically, the scale-free property implies preferential attachment [22, 49]. In addition, from Fig. 1(b), one can find that the predictive accuracies of these PA indices are generally higher in the disassortative networks than those in the assortative ones, although there is not an obvious and direct correlation between assortative coefficient and algorithmic accuracies based on these PA indices.

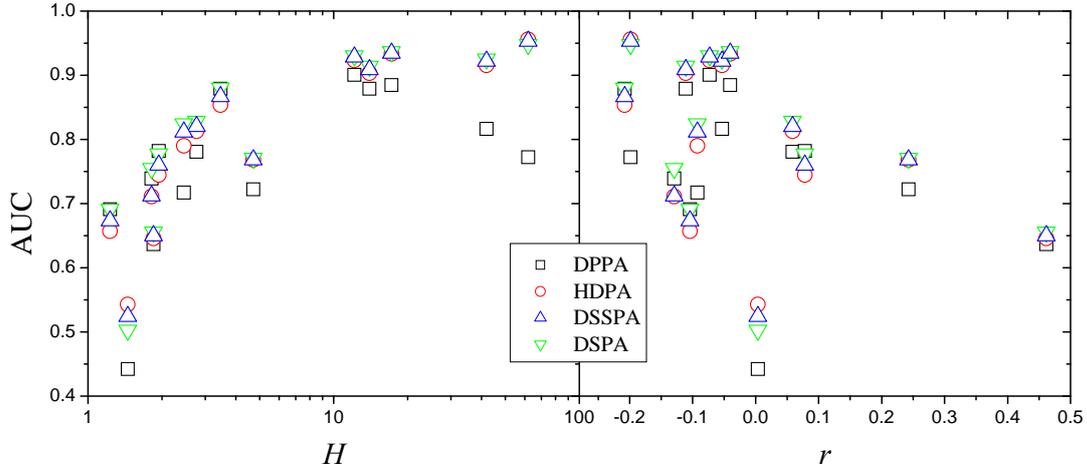

Fig. 1: (Color online) AUC accuracies of these PA indices versus (a) the degree heterogeneity H, (b) the degree-degree correlation coefficient.

On the other hand, in the same network, their accuracies are very different as well. Intuitively, the DPPA index may be most reasonable for link prediction among the above three indices. However, the results in table 2 show that the DSPA and DSSPA indices clearly outperform the DPPA index. For example, in the INT, WWW and GP, the accuracies of DSPA are much higher than DPPA about 22 percent. In addition, the HDPA index is very simple and depends on the degree of single node, but it is still effective and even better than the DPPA index sometimes. According to the definitions of the these PA indices, significant difference about similarity scores does not lies in the links between pairs of high-degree nodes since these links are assigned high scores by all these indices. However, for all other links, the DPPA index usually endows the assortative links (i.e., the links between pairs of the same degree level) with relatively high scores, while the HDPA, DSPA and DSSPA indices can give competitive scores for both the disassortive links (i.e., links between high-degree nodes and low-degree nodes) and the assortative links between high-degree nodes. For example, we assume that the two nodes connected by an assortative link $L_{ij}$ have the same degrees of $k_i=k_j=20$; and the two nodes connected by a disassortative link $L_{i'j'}$ have very different degrees, $k_{i'}=2$ and $k_{j'}=100$. Obviously, the score of $L_{ij}$ is higher than that of $L_{i'j'}$ in the DPPA index, while in the HDPA, DSPA and DSSPA indices, it is lower than that of $L_{i'j'}$. Since many real networks possess scale-free property, there are a large number of links between high-degree nodes and other nodes. Especially in some networks with the strong heterogeneity and disassortative correlations, the probability that a link exists between high-degree nodes and other nodes may be larger than the probability between pairs of medium-degree nodes. Thus, if the investigated network simultaneously has large degree heterogeneity and disassortative correlation, such as the INT, WWW, GP, EE and ES, both the HDPA, DSPA and DSSPA indices perform better than the DPPA index. Moreover, the effect of the heterogeneity is relatively more remarkable than the disassortative correlation. In general, higher is the heterogeneity of a network, higher is the accuracies of the HDPA, DSPA and DSSPA indices. For instance, the INT has extremely large degree heterogeneity, so the performances of the HDPA, DSPA and DSSPA algorithms (AUC can achieve about 0.95) are remarkably better than that of the DPPA one (AUC is about

0.77).

From the perspective of the rules of network evolution, the results in table 2 are also intuitively reasonable. If a mechanism of link formation can properly model the link formation in real complex networks, it can be considered as a kind of link prediction algorithm; conversely a high-accuracy predictive algorithm may suggest a possible mechanism to dominate a network evolution [6]. As shown in table 2, the HDPA index has good performance and even gives higher accuracies than the DPPA one in some networks. This indicates that the HDPA index indeed captures the preferential attachment mechanism to some extent, and on the other hand, the effectiveness of the HDPA index also give a indirect evidence that the networks is organized under the PA mechanism. In addition, among the DPPA, DSSPA and DSPA indices, the DSPA index has the best performance on most of the sample networks. According to the analysis mentioned in section 2.2, the DSPA index can be viewed as a special case of the general similarity index for internal links by Eq. (5), while the DPPA and DSSPA indices are two limiting cases of the GILPA one. Thus we think the DSPA index can more completely reflect the mechanism of internal link formation than the DPPA and DSSPA ones.

Finally, it should be pointed out that the degree-summation form may be the simplest form for the general similarity index given by Eq. (5), but perhaps it is not one with the highest accuracy. Under this consideration, we also calculate the accuracies of the general similarity index with different parameter $\varepsilon$, and find the optimal values of $\varepsilon$ for the fourteen networks, which is presented in last column in Table. 1. Interestingly, the highest accuracies are not significantly higher than, even are equal to, that of the DSPA index. Perhaps, the mechanism of addition of internal links is indeed follows the degree summation form.

## 4 Conclusions

In summary, based on the popular network evolution mechanism, i.e., the PA mechanism, we have developed a class of new PA similarity indices to estimate the likelihood of the existence of a link between two nodes. By applying them to fourteen real networks, we have shown that the proposed indices can provide more accurate predictions than the traditional DPPA index, especially in the networks with the large degree heterogeneity and the disassortative degree correlation. Moreover, the computational complexity of these indices is almost the same as or lower than the DPPA index, we believe, they can provide competitively effective and efficient prediction as the DPPA index. In addition, owing to the important correlation between link-prediction algorithm and mechanism of network evolution [6], we hope that this work is helpful to understand the mechanism of network evolution, especially for the formation mechanism of the internal links.

Of course, in this paper, we do not consider the link deletion although it is an elementary process for network evolution [24]. Moreover, according to our assumption, for the links from new nodes, we should apply the HDPA index, while for the internal links, the DSPA index may be more suitable. But it is usually difficult to distinguish the two types of links in most real networks. Thus we do not know which index should be better to perform the link prediction. Moreover, the degrees of new nodes often are very small, and except for the link formation from new nodes to old nodes, there may also exists the link formation from the old nodes to the new nodes in some networks when the new nodes are added into networks. Under this consideration, one should be advised to use the DSPA index.

Finally, we also note that the weight of link in the weighted networks has been recently introduced into the problem of link prediction and significant improvements about link-prediction accuracies are presented [19]. By replacing the node's degree with node's strength, these PA indices can be also easily extended to the corresponding weighted versions. We hope that in weighted networks, the introduction of weight can further improve the performance of these PA algorithms, which will be investigated in our future works.

**Acknowledgement**

This work has been supported by the National Natural Science Foundation of China (Grant Nos. 6104104, 11147121, and 61104143), the Scientific Research Fund of Education Department of Hunan Province of China (Grant No. 11B128), and partly by the Doctor Startup Project of Xiangtan University (Grant No. 10QDZ20).

**References:**
[1]  L. Getoor, C.P. Diehl, ACM SIGKDD Explorations Newsletter 7 (2005) 3-12.
[2]  L. Lü, T. Zhou, Physica A 390 (2011) 1150-1170.
[3]  A. Clauset, C. Moore, M.E.J. Newman, Nature 453 (2008) 98-101.
[4]  S. Redner, Nature 453 (2008) 47-48.
[5]  R. Guimerà, M. Sales-Pardo, Proc. Natl. Acad. Sci. U.S.A. 106 (2009) 22073-22078.
[6]  W.-Q. Wang, Q.-M. Zhang, T. Zhou, Europhys. Lett. 98 (2012) 28004.
[7]  H.-K. Liu, L. Lü, T. Zhou, Sci. China Ser. G 41 (2011) 816-823.
[8]  H. Yu, P. Braun, M.A. Yildirim, etal, Science 322 (2008) 104-110.
[9]  M.P.H. Stumpf, T. Thorne, E. de Silva, R. Stewart, H.J. An, M. Lappe, C. Wiuf, Proc. Natl. Acad. Sci. U.S.A. 105 (2008) 6959-6964.
[10]  L.A.N. Amaral, Proc. Natl. Acad. Sci. U.S.A. 105 (2008) 6795-6796.
[11]  B. Gallagher, H. Tong, T. Eliassi-Rad, C. Faloutsos, Proceedings of the ACM SIGKDD International Conference on Knowledge Discovery and Data Mining (ACM Press) 2008, pp. 256–264.
[12]  Q.-M. Zhang, M.-S. Shang, L. Lü, Int. J. Mod. Phys. C 21 (2010) 813-824.
[13]  R.R. Sarukkai, Computer Networks 33 (2000) 377-386.
[14]  J. Zhu, J. Hong, J.-G. Hughes, in: Proceedings of the First International Conference on Computing in an Imperfect World, 8-10 April 2002, p.60-73.
[15]  A. Popescul, L.H. Ungar, in: IJCAI Workshop on Learning Statistical Models from Relational Data, ACM Press, New York, 2003.
[16]  K. Yu, W. Chu, S. Yu, V. Tresp, Z. Xu, Advance in Neural Information Processing Systems 19 (2007) 1553-1560.
[17]  T. Zhou, L. LÄu, Y.-C. Zhang, Eur. Phys. J. B 71 (2009) 623-630.
[18]  L. Lü, C.-H. Jin, T. Zhou, Phys. Rev. E 80 (2009) 046122.
[19]  L. Lü, T. Zhou, Europhys. Lett. 89 (2010) 18001.
[20]  W.-P. Liu, L. Lü, Europhys. Lett. 89 (2010) 58007.
[21]  Z. Liu, Q.-M. Zhang, L. Lü, T. Zhou, Europhys. Lett. 96 (2011) 48007.
[22]  A.L. Barabási, R. Albert, Science 286 (1999) 509-511.
[23]  S.N. Dorogovtsev, J.F.F. Mendes, Europhys. Lett. 52 (2000) 33-39.
[24]  R. Albert, A.L. Barabási, Phys. Rev. Lett. 85 (2000) 5234–5237.
[25]  W.-X. Wang, B.-H. Wang, B. Hu, G. Yan, Q. Ou, Phys. Rev. Lett. 94 (2005) 188702.
[26]  P.L. Krapivsky, S. Redner, Computer Networks 39 (2002) 261–276.
[27]  V. Rosato, F. Tiriticco, Europhys. Lett. 66 (2004) 471-477.
[28]  B. Tadic, Physica A 293, 273–284 (2001); Physica A 314 (2002) 278–283.

[29] A.L. Barabási, H. Jeong, Z. Néda, E. Ravasz, A. Schubert, T. Vicsek, Physica A 311 (2002) 590-614.

[30] V. Batagelj, M. Zaveršnik, Pajek Datasets, available at http://vlado.fmf.uni-lj.si/pub/networks/data/collab/geom.htm

[31] This snapshot was created by Newman M E J from data for July 22, 2006 and is not previously published, available at http://www-personal.umich.edu/~mejn/netdata/

[32] V. Batagelj, A. Mrvar, Pajek Datasets, available at http://vlado.fmf.uni-lj.si/pub/networks/data/default.htm

[33] D.B. Bu, Y. Zhao, L. Cai, H. Xue, X.P. Zhu, H.C. Lu, J.F. Zhang, S.W. Sun, L.J. Ling, N. Zhang, G.J. Li, R.S. Chen, Nucleic Acids Research 31 (2003) 2443-2450.

[34] C.J. Melián, J. Bascompte, Ecology 85 (2004) 352-358.

[35] M.E.J. Newman, Phys. Rev. E 74 (2006) 036104.

[36] R. Albert, H. Jeong, A.-L. Barabási, Nature 401 (1999) 130-131.

[37] M. Ripeanu, I. Foster, A. Iamnitchi, IEEE Internet Computing Journal 2429 (2002) 85-93.

[38] J. Leskovec, J. Kleinberg, C. Faloutsos, in: Proceedings of the ACM SIGKDD international conference on knowledge discovery and data mining, Chicago, 21-24 August 2005, p.177-187.

[39] M. Richardson, R. Agrawal, P. Domingos, in: Proceedings of the Second International Semantic Web Conference, 2003, p.351-368.

[40] J. Leskovec, K. Lang, A. Dasgupta, M. Mahoney, Community Structure in Large Networks: Natural Cluster Sizes and the Absence of Large Well-Defined Clusters, Preprint, 2008, arXiv:0810.1355v1[cs.DS].

[41] R. Guimerà, L. Danon, A. Díaz-Guilera, F. Giralt, A. Arenas, Phys. Rev. E 68 (2003) 065103(R).

[42] D. J. Watts and S. H. Strogatz, Nature 393 (1998) 440-442.

[43] V. Latora, M. Marchiori, Phys. Rev. Lett. 87 (2001) 198701.

[44] M.E.J. Newman, Phys. Rev. Lett. 89 (2002) 208701.

[45] P. Holme, B.J. Kim, C.N. Yoon, S.K. Han, Phys. Rev. E 65 (2002) 056109

[46] C.-Y. Yin, W.-X. Wang, G.-R. Chen, B.-H. Wang, Phys. Rev. E 74 (2006) 047102.

[47] G.-Q. Zhang, D. Wang, G.-J. Li, Phys. Rev. E 76 (2007) 017101.

[48] J.A. Hanely, B.J. McNeil, Radiology 143 (1982) 29-36.

[49] K.A. Eriksen, M. Hörnquist, Phys. Rev. E 65 (2001) 017102.